\documentstyle[12pt]{article}

\def \beq { \begin{equation} }
\def \eeq { \end{equation} }

\def \0j {j_{{}_0} }
\def \1j {j_{{}_1} }

\begin{document}

\title{Algebra of non-local charges in the O(N) WZNW model 
at and beyond criticality}

\author{L.E. Saltini\thanks{Work supported by FAPESP. E-mail:
lsaltini@fma1.if.usp.br}\\
Instituto de F\'\i sica da Universidade de S\~ao Paulo\\
C.P. 66318, S\~ao Paulo SP, Brazil, 05315-970\\
{}\\
A. Zadra\thanks{Work supported by FAPESP. E-mail: 
zadra@crm.umontreal.ca}\\
Centre de Recherches Math\'ematiques, Universit\'e de Montr\'eal\\
C.P. 6128-A, Montr\'eal QC, Canada, H3C 3J7}

\date{}

\maketitle

\begin{abstract}
We derive the classical algebra of the non-local conserved charges 
in the $O(N)$ WZNW model and analyze its dependence on the 
coupling constant of the Wess-Zumino term. As in the non-linear 
sigma model, we find cubic deformations of the $O(N)$ 
affine algebra. The surprising result is that the cubic algebra of 
the WZNW non-local charges does not obey the Jacobi identity, 
thus opposing our expectations from the known Yangian symmetry of 
this model.

\end{abstract}

\section{ Introduction}

Yangian symmetries are expected to play a major role in our 
understanding of the integrable structure of conformal field 
theories and their deformations \cite{1,2}. Some conformal field 
theories are known to exhibit a Yangian symmetry for any affine 
Lie algebra at the critical point, with a level-independent 
structure \cite{3,4,5}. The Yangian generators of that symmetry 
are understood as quantum extensions of classical non-local 
charges, such as those found in the non-linear sigma model and 
current algebra models \cite{6}-\cite{15}. Therefore the study 
of classical algebras of non-local charges may be regarded as 
a pre-quantum step toward the comprehension of symmetry and 
integrability properties of this class of field theories.

In previous works \cite{16,17}, we have studied the algebra 
of the infinite non-local conserved charges in the $O(N)$ 
non-linear sigma model. The Wess-Zumino-Novikov-Witten (WZNW) 
model is also known to display an infinite set of non-local 
charges \cite{18}, and the primary aim of this paper is to 
unveil the classical algebra generated by them. 
In particular, we are interested in the dependence of this 
algebra with respect to the Wess-Zumino coupling constant. 
By understanding this dependence we could approach the algebra 
simultaneously at and beyond the conformal point, defined by 
a specific value of that coupling. Therefore, 
one of the applications -- and the main motivation -- of our algebraic  
project is the study of integrable perturbations of conformal theories 
\cite{19,20,21}. 

To construct the charges and the corresponding Dirac brackets, 
we follow the algebraically inspired strategy outlined in 
ref.\cite{16} and the diagrammatic technique introduced in ref.\cite{17}. 
As a result, we observe, just like in the previously studied theories 
(namely, the bosonic and supersymmetric non-linear sigma models and the 
Gross-Neveu model), that the WZNW non-local charges turn out to obey a 
cubic deformation of the $O(N)$ affine algebra. Those cubic algebras 
are not isomorphic though. Surprisingly, the algebra of the WZNW charges 
does not satisfy the Jacobi identity, as opposed to the algebra of 
the chiral non-linear sigma model. These are the main results to be 
reported in this letter.

In the next section, we take the $O(N)$ WZNW model, recall its 
current algebra and related properties, and define some 
notation. In section 3, we discuss the available methods that can be used 
to construct the non-local charges. In section 4, we consider the WZNW 
in the conformal point and derive the corresponding algebra of non-local 
charges. Then we study the same algebra 
for the WZNW model beyond the conformal point, in section 5, 
including a discussion on the zero-coupling limit. 
In section 6 we analyze the issue of the Jacobi identity for the cubic 
algebra. Section 7 is used to summarize results and comments.

\section{ Current algebra in the WZNW model}

Our starting point is the well-known WZNW action which contains two parcels, 
\beq 
S=S_{ch}+n S_{WZ} \quad ;
\label{action}
\eeq
$S_{ch}$ is the action of the ordinary chiral model,
\beq
S_{ch}=-\frac{1}{2\lambda ^2}\int d^{2}x
\eta^{\mu\nu}tr(g^{-1}\partial_{\mu}g g^{-1}\partial_{\nu}g) \,\,\, ,
\eeq
where the basic field $g(x)$ takes values 
in a simple Lie group $G$ (we shall take 
$G=O(N)$), $n$ is an integer and $S_{WZ}$ is the Wess-Zumino term
\beq
S_{WZ}=\frac{1}{4\pi}\int_{0}^{1}dr\int d^{2}x \epsilon^{\mu\nu} 
tr(\tilde{g}^{-1}\partial_{r}\tilde{g}\tilde{g}^{-1}\partial_{\mu}\tilde{g}
\tilde{g}^{-1}\partial_{\nu}\tilde{g}) \,\,\, .
\eeq
(Here, $B$ is an appropriate three-dimensional boundary of $G$ and 
the extended field $\tilde g$ is assumed to be constant outside 
a tubular neighborhood $\Sigma \times [0,1]$ of the boundary $\Sigma $ of 
$B$; $r$ is the coordinate normal to the boundary.) This model contains a 
free coupling constant $\lambda $ such that, for $\lambda ^2 \to 0$, 
we recover the ordinary chiral model and, for $\lambda ^2= 4\pi /n$, 
the conformally invariant WZNW model.

The action (\ref{action}) has a global invariance under the product 
group $G_L \times G_R$, which acts on $G$ according to
\beq
g\to g_L g g_R^{-1} \quad .
\eeq
This invariance leads to conserved Noether currents, taking values in the 
Lie algebra ${\cal G}_L \oplus {\cal G}_R$, which may be decomposed as 
left and right currents. For the purpose of writing a current algebra, it 
is convenient to work with covariant currents. To do so, one may begin with 
the following left and right non-conserved currents,
\beq
j^{L}_{\mu}=-\frac{1}{\lambda^{2}}\partial_{\mu}g g^{-1} \,\,\, ,
\,\,\, j^{R}_{\mu}=+\frac{1}{\lambda^{2}}g^{-1}\partial_{\mu}g \,\,\, ,
\eeq
whose algebra, for a general simple Lie group, 
was derived in ref.\cite{19}. Here we shall limit ourselves to the $O(N)$ 
algebra. We shall also introduce the parameter
\beq
\alpha = {n\lambda ^2\over 4\pi} \quad 
\eeq
which appears in the definition of the conserved covariant currents 
\begin{eqnarray}
J^{L}_{\mu}=(\eta_{\mu\nu}+\alpha\epsilon_{\mu\nu})j^{L\nu}&=& 
-\frac{1}{\lambda^{2}}
(\eta_{\mu\nu}+\alpha\epsilon_{\mu\nu})\partial^{\nu}g g^{-1} \,\,\, ,
\nonumber \\
J^{R}_{\mu}=(\eta_{\mu\nu}-\alpha\epsilon_{\mu\nu})j^{R\nu}&=& 
+\frac{1}{\lambda^{2}}
(\eta_{\mu\nu}-\alpha\epsilon_{\mu\nu})g^{-1}\partial^{\nu}g  \,\,\, .
\label{newcu}
\end{eqnarray}
It is quite convenient to use the $\circ $-product notation 
introduced in \cite{16}, which is a characteristic structure the 
$O(N)$ algebra,
\beq
(A \circ B)_{ij,kl}\equiv 
A_{ik}B_{jl}-A_{il}B_{jk}+A_{jl}B_{ik}-A_{jk}B_{il} \,\,\, ,
\eeq
to write down the classical current algebra \cite{19} 
in the following way:
\begin{eqnarray}
\{(J^{L}_{0})_{ij}(x),(J^{L}_{0})_{kl}(y)\}&=&(I \circ
J^{L}_{0})_{ij,kl}(x) \delta(x-y)-\alpha(I \circ I)_{ij,kl}
\delta'(x-y) \,\,\, , \nonumber \\
\{(J^{L}_{0})_{ij}(x),(J^{L}_{1})_{kl}(y)\}&=&(I \circ
J^{L}_{1})_{ij,kl}(x) \delta(x-y)-\frac{(1+\alpha^{2})}{2}(I \circ I)_{ij,kl}
\delta'(x-y) \,\,\, , \nonumber \\
\{(J^{L}_{1})_{ij}(x),(J^{L}_{1})_{kl}(y)\}&=&2\alpha(I \circ
J^{L}_{1})_{ij,kl}(x) \delta(x-y)-\alpha^{2}(I \circ
J^{L}_{0})_{ij,kl}(x) \delta(x-y) \nonumber \\
&-&\alpha(I \circ I)_{ij,kl}\delta'(x-y) \,\,\, , \nonumber 
\end{eqnarray}

\begin{eqnarray}
\{(J^{R}_{0})_{ij}(x),(J^{R}_{0})_{kl}(y)\}&=&(I \circ
J^{R}_{0})_{ij,kl}(x) \delta(x-y)+\alpha(I \circ I)_{ij,kl}
\delta'(x-y) \,\,\, , \nonumber \\
\{(J^{R}_{0})_{ij}(x),(J^{R}_{1})_{kl}(y)\}&=&(I \circ
J^{R}_{1})_{ij,kl}(x) \delta(x-y)-\frac{(1+\alpha^{2})}{2}(I \circ I)_{ij,kl}
\delta'(x-y) \,\,\, , \nonumber \\
\{(J^{R}_{1})_{ij}(x),(J^{R}_{1})_{kl}(y)\}&=&-2\alpha(I \circ
J^{R}_{1})_{ij,kl}(x) \delta(x-y)-\alpha^{2}(I \circ
J^{L}_{0})_{ij,kl}(x) \delta(x-y) \nonumber \\
&+&\alpha(I \circ I)_{ij,kl}\delta'(x-y) \,\,\, , \nonumber 
\end{eqnarray}

\begin{eqnarray}
\{(J^{L}_{0})_{ij}(x),(J^{R}_{0})_{kl}(y)\}&=&0 \,\,\, , \nonumber \\
\{(J^{L}_{0})_{ij}(x),(J^{R}_{1})_{kl}(y)\}&=&
-(1-\alpha^{2})(g \circ g)_{ij,kl}(y)\delta'(x-y) \,\,\, , \nonumber \\
\{(J^{L}_{1})_{ij}(x),(J^{R}_{0})_{kl}(y)\}&=&
-(1-\alpha^{2})(g \circ g)_{ij,kl}(x)\delta'(x-y) \,\,\, , \nonumber \\
\{(J^{L}_{1})_{ij}(x),(J^{R}_{1})_{kl}(y)\}&=&
-\alpha(1-\alpha^{2})(g \circ g)_{ij,kl}(y)\delta'(x-y) \nonumber \\ 
&+&\alpha(1-\alpha^{2})(g \circ g)_{ij,kl}(x)\delta'(x-y)\,\,\, . \nonumber 
\end{eqnarray}
This algebra is invariant under the combined change $L\leftrightarrow R$ and 
$\alpha \leftrightarrow -\alpha$. Therefore we may concentrate ourselves in 
one sector and easily extend the results to the other. Let us notice the 
Schwinger terms, both in time and space-component brackets, 
whose present form we believe to be underneath the unexpected 
properties of the algebras to be shown in this paper.

\section{ Non-local charges}

One can easily check that the conserved covariant currents ($J^{R,L}_\mu$)  
satisfy a curvature-free condition,
\beq
\partial_{\mu}J^{R,L}_{\nu}-\partial_{\nu}J^{R,L}_{\mu}+
\lambda^{2}[J^{R,L}_{\mu},J^{R,L}_{\nu}]=0 \,\,\, ,
\eeq
which implies the existence of an infinite set of non-local 
conserved charges, in both left and right sectors. One could then use the 
integro-differential algorithm of Br\'ezin {\it et.al.} \cite{8} to 
construct an infinite set of non-local conserved charges 
$Q^{(n)}, n=0,1,\cdots $. Besides the (local) $O(N)$ 
generator
\beq
(Q^{(0)})_{ij}=\int dx\, (J_{0})_{ij} \,\,\, ,
\eeq
we also recall the standard expression of the first non-local charge 
\cite{18} 
\beq
(Q^{(1)})_{ij}=\int dx\, (J_{1}-\alpha J_0+2J_{0}\partial^{-1}J_{0})
_{ij} \,\,\, .
\eeq  
However, from the algebraic point of view, the set of charges thus 
generated is not necessarily the most suitable. In ref.\cite{16}, after 
studying the non-linear sigma model, it was shown that the standard 
charges from Br\'ezin {\it et.al.}'s algorithm can be recombined 
into a new set of improved charges, whose algebra is somewhat 
simpler -- it still is a non-linear algebra but the non-linear terms 
are simply cubic. Taking this ``algebraic simplicity" as a guiding 
criterion, the remaining 
improved charges are constructed from the algebra itself, using the 
charge $Q^{(1)}$ as a step-like generator. This procedure is made 
possible due to the property
\beq
Q^{(n+1)}\propto \,\,\, linear \,\,\,\, part \,\,\,\, of \,\,\,\,
\{Q^{(n)},Q^{(1)}\}  \,\,\, .
\eeq   

Therefore our task is to apply a similar algebraic procedure to the 
WZNW model and to find the corresponding set of improved non-local 
charges, whose algebra is supposedly as simple 
as possible. The calculations involved in this program can be shortened 
if we use the diagrammatic method developed in ref.\cite{17}. 
It consists of 
a graphic representation of charges and brackets, which incorporates  
currents, non-localities and contraction rules in a rather handy way. The 
graphic rules from ref.\cite{17} can be adapted to the WZNW model with no 
major difficulty. The results are discussed in the next sections.

\section{ Charges and algebra at the critical point}

The critical value of the coupling constant, for which the model is 
conformally invariant, corresponds to $\alpha = \pm 1$. In that case, 
the covariant current components are chirally constrained
\beq
J_0^L=J_1^L \qquad ,\qquad J_0^R=-J_1^R
\eeq
and the non-local covariant charges can be written in terms of the 
time-component 
$J_0$ solely,
\begin{eqnarray}
Q^{(0)} &=& \int dx \, (J_0)  \quad ,\nonumber \\
Q^{(1)} &=& \int dx \, J_0 2\partial ^{-1} (J_0)\quad ,\nonumber \\
&\vdots & \nonumber \\
Q^{(n)} &=& \int dx \, J_0 2\partial ^{-1} (J_0 
2\partial ^{-1} (J_0  2\partial ^{-1} (J_0 2\partial ^{-1} \cdots )))
\quad .
\end{eqnarray}
According to the terminology proposed in ref.\cite{16}, we would say  
the critical charges are ``saturated", which just means that each non-local 
charge is made out of a single chain of time-components $J_0$'s 
connected by the non-local operator $\partial ^{-1}$. 
The algebra therefore depends on $\{J_0 ,J_0\} $ only and is readily 
derived using the graphic method. The resulting cubic algebra is  
briefly presented in terms of generators, as follows:
\beq
\{Q(\xi),Q(\mu)\}=\left( f(\xi,\mu) \circ Q(\xi)-Q(\mu) \right) \quad ,
\label{critalg}
\eeq
where $\xi$ and $\mu $ are expansion parameters in the charge-generator 
matrix defined by
\beq
Q(\xi) \equiv \sum_{n=0}^{\infty}\xi^{n+1}Q^{(n)}
\eeq
and $f$ is a two-parameter dependent matrix,
\beq
f(\xi ,\mu ) = \frac{1-2\alpha(\xi+\mu)}{\xi^{-1}-\mu^{-1}}
\left( I- Q(\xi)Q(\mu) \right) \quad .
\label{fcrit}
\eeq
In this formula, $I$ is the $N\times N$ identity matrix which leads to 
the linear part of the algebra. On the other hand, the quadratic term 
$Q(\xi)Q(\mu)$ in $f$ implies the cubic piece in that same algebra. The 
linear part was derived in full generality, using 
the graphic method, while the cubic terms were verified up to the order
$n=4$ (thus the quadratic part in (\ref{fcrit}) should be regarded 
as an ansatz).

We recall that $\alpha = \pm 1$ at criticality and the results above are 
understood to hold only in those cases. Yet it is worth noticing that, 
should we simply set $\alpha =0$, the resulting algebra would be 
isomorphic to the cubic algebra of the non-linear sigma model \cite{16,17}. 
However, as we show in section 5, the cases 
$\alpha \not =\pm 1$ display an authentically new algebra.

\section{ Non-local charges beyond the critical point}

In the non-linear sigma model, the following coincidence was 
observed: the algebras of the improved charges and of 
the (non-conserved) saturated charges are identical \cite{16}. 
It was conjectured in ref.\cite{16} 
that the same property might hold for the WZNW model, i.e. that the 
algebra (\ref{critalg}) would be obeyed for any value of the 
coupling $\alpha $. However, the following results imply that that 
conjecture is false: we did find a cubic algebra, and some of its 
brackets were listed below,

\begin{eqnarray}
\{Q^{(0)},Q^{(0)}\}&=&(I \circ Q^{(0)}) \,\,\, ,
\nonumber \\
\{Q^{(0)},Q^{(1)}\}&=&(I \circ Q^{(1)}) 
-2\alpha (I \circ Q^{(0)}) \,\,\, , \nonumber \\
\{Q^{(1)},Q^{(1)}\}&=&(I \circ Q^{(2)}) 
-4\alpha (I \circ Q^{(1)})-(Q^{(0)}Q^{(0)}\circ Q^{(0)}) 
\,\,\, , \nonumber \\
\{Q^{(0)},Q^{(2)}\}&=&(I \circ Q^{(2)}) 
-2\alpha (I \circ Q^{(1)}) 
-(1-\alpha^{2})(I \circ Q^{(0)})\,\,\, , \nonumber \\
\{Q^{(1)},Q^{(2)}\}&=&(I \circ Q^{(3)}) 
-4\alpha (I \circ Q^{(2)})-(Q^{(1)}Q^{(0)}\circ Q^{(0)})- \nonumber \\
&-&(Q^{(0)}Q^{(0)}\circ Q^{(1)})+
2\alpha(Q^{(0)}Q^{(0)}\circ Q^{(0)})\,\,\, , \nonumber \\
\{Q^{(0)},Q^{(3)}\}&=&(I \circ Q^{(3)}) 
-2\alpha (I \circ Q^{(2)})-(1-\alpha^{2})(I \circ
Q^{(1)}) - \nonumber \\
&-&\alpha(1-\alpha^{2})(I \circ Q^{(0)}) \,\,\, . \nonumber \\
\{Q^{(1)},Q^{(3)}\}&=&(I \circ Q^{(4)}) 
-4\alpha (I \circ Q^{(3)})-(Q^{(2)}Q^{(0)}\circ Q^{(0)})-\nonumber \\
&-&(Q^{(1)}Q^{(0)}\circ Q^{(1)})
-(Q^{(0)}Q^{(0)}\circ Q^{(2)})+ \nonumber \\
&+&2\alpha (Q^{(1)}Q^{(0)}\circ Q^{(0)})
+2\alpha (Q^{(0)}Q^{(0)}\circ Q^{(1)})+ \nonumber \\
&+&2(1-\alpha^{2})(Q^{(0)}Q^{(0)}\circ Q^{(0)})\,\,\, , 
\nonumber \\
\{Q^{(2)},Q^{(2)}\}&=&(I \circ Q^{(4)}) 
-4\alpha (I \circ Q^{(3)}) 
-3\alpha(1-\alpha^{2})(I \circ Q^{(1)})-\nonumber \\ 
&-&3\alpha^{2}(1-\alpha^{2})(I \circ Q^{(0)})
-(Q^{(0)}Q^{(0)}\circ Q^{(2)})+\nonumber \\ 
&+&2\alpha(Q^{(0)}Q^{(1)}\circ Q^{(0)})  
+2\alpha(Q^{(1)}Q^{(0)}\circ Q^{(0)})- \nonumber \\
&-&(Q^{(0)}Q^{(1)}\circ Q^{(1)})
-(Q^{(1)}Q^{(0)}\circ Q^{(1)})- \nonumber \\
&-&(Q^{(1)}Q^{(1)}\circ Q^{(0)})
+4\alpha(Q^{(0)}Q^{(0)}\circ Q^{(1)})\,\,\, ,
\label{ncritalg}
\end{eqnarray}
but we found no linear change of basis such that it might turn back into 
the algebra (\ref{critalg}) for any $\alpha $. Moreover, if this algebra 
could really be written in a form similar to (\ref{critalg}), 
then $f(\xi ,\mu )$ would not be given by an expression as simple as 
(\ref{fcrit}). So far, we do not have an ansatz for the complete algebra. 
Nevertheless, if necessary, other charges and brackets can be generated 
from (\ref{ncritalg}), using the graphic algorithm.

The zero-coupling case ($\alpha \to 0$) was studied separately 
because, in that limit, we expected to reach an algebra isomorphic  
to the one found in the non-linear sigma model. 
We constructed some non-local charges and calculated various brackets, 
up to $\{Q^{(4)},Q^{(2)}\}$, and all brackets thus found fit into the 
following ansatz:
\beq
\{Q(\xi ),Q(\mu )\} = (f(\xi, \mu)\circ c(\xi ,\mu)Q(\xi ) -
c(\mu ,\xi)Q(\mu ))\quad ,
\eeq
\beq
Q(\xi )=\sum _{n=0}^\infty \xi^{n+1}Q^{(n)} \quad ,\quad
f(\xi ,\mu) ={1\over \xi^{-1} -\mu ^{-1}}(I-Q(\xi )Q(\mu )) \quad ,
\eeq
\beq
c(\xi ,\mu) = 1 - \xi^2 + 2\xi^4 -\xi^2 \mu^2 + \cdots
\eeq
Further terms in the expansion of the function $c(\xi, \mu)$ 
would require higher-order brackets. As is turns out, the 
above algebra is different from the one of the non-linear 
sigma model \cite{16,17}. This divergence is directly related to 
the differences between the Schwinger terms in the 
respective current algebras.
 
In the search for a general ansatz, we asked ourselves 
whether we could use the Jacobi identity to derive a 
closed form for the $\alpha$-dependent cubic algebra. 
This actually happens in the non-linear 
sigma model: from the Jacobi identity with low-order charges one can 
calculate higher-order brackets. 
This procedure is discussed in the next section.

\section{ On the Jacobi identity}

Based on the available algebras, 
we posed ourselves the following question: 
assuming a cubic algebra with the structure
\begin{eqnarray}
& &\{ Q(\xi ),Q(\mu ) \} =(f(\xi,\mu) \circ Q(\xi)-Q(\mu))\quad ,\\
& &f(\xi,\mu)=A(\xi,\mu)I+B(\xi,\mu)Q(\xi)Q(\mu)\quad ,
\end{eqnarray}
what constraints would the Jacobi identity impose on the  
to-be-determined functions $A(\xi ,\mu )$ and $B(\xi ,\mu )$ ? 
To begin with, we considered $A$ and $B$ to be ordinary two-parameter 
functions, in which case the answer is
\begin{eqnarray}
& & A(\xi ,\mu ) = \frac{1}{g(\xi)-g(\mu)} \label{A}\quad ,\\
& & B(\xi ,\mu ) = {\rm constant} \times A(\xi ,\mu )\quad ,
\end{eqnarray}
where $g(\xi)$ is some arbitrary function. This solution is 
compatible with the non-linear sigma model, where it was found 
$A=-B=1/(\xi^{-1} -\mu^{-1})$. Actually that solution is rather
unique: (i) Firstly, by an overall rescaling of 
the generator $Q(\xi)$, we could set the constant = -1 in the solution
above, so that $A=-B$ could be taken as a general relation; 
(ii) As long as $g(\xi)$ is invertible and if the inverse
$g^{-1}$ admits a series expansion, we might take $\xi = g^{-1} (\xi ')$ 
and define another generator $Q'(\xi ')=Q(g^{-1}(\xi '))= \sum \xi '
Q^{\prime (n)}$, where $Q^{\prime (n)}$ would be linear recombinations 
of the original non-local charges. In the re-parameterized basis, we 
would reproduce the cubic algebra of the non-linear sigma 
model.

However, the critical ($\alpha =1$) WZNW model has 
\beq
A=-B= {1-2(\xi +\mu)\over \xi ^{-1} - \mu ^{-1}} \quad ,
\eeq
which cannot be written in the form (\ref{A}) 
and therefore the cubic algebra (\ref{critalg}) does 
not satisfy the Jacobi identity, as announced in the Introduction. 
It is important to note that the breaking of the Jacobi identity 
is not caused by the cubic terms; in fact, the linear part of 
the algebra is sufficient to break it. The following test, 
taken from the $\alpha =1$ model, exemplifies this property:
\begin{eqnarray}
& &\{\{Q^{(0)}_{ij},Q^{(1)}_{kl}\},Q^{(1)}_{mn}\}+
\{\{Q^{(1)}_{kl},Q^{(1)}_{mn}\},Q^{(0)}_{ij}\}+
\{\{Q^{(1)}_{mn},Q^{(0)}_{ij}\},Q^{(1)}_{kl}\}= \nonumber \\
&=& 4\left[ (\delta_{ik}\delta_{lm}-\delta_{il}\delta_{km})Q^{(0)}_{jn}+
(\delta_{jl}\delta_{km}-\delta_{jk}\delta_{lm})Q^{(0)}_{in}+
\right. \nonumber \\
& &\quad 
+(\delta_{jm}\delta_{kn}-\delta_{km}\delta_{jn})Q^{(0)}_{il}+
(\delta_{in}\delta_{km}-\delta_{im}\delta_{kn})Q^{(0)}_{jl}+ 
\nonumber \\
& &\quad 
+(\delta_{il}\delta_{kn}-\delta_{ik}\delta_{ln})Q^{(0)}_{jm}+
(\delta_{jk}\delta_{ln}-\delta_{jl}\delta_{kn})Q^{(0)}_{im}+
\nonumber \\
& &\quad 
+\left. (\delta_{jn}\delta_{lm}-\delta_{jm}\delta_{ln})Q^{(0)}_{ik}+
(\delta_{ln}\delta_{im}-\delta_{in}\delta_{lm})Q^{(0)}_{jk} \right]
\quad .
\end{eqnarray}

Although a general ansatz for the off-critical algebra is missing, 
we have also made tests with the available brackets, such as this:
\begin{eqnarray}
& &linear\; part\; of\; \{\{Q^{(1)}_{ij},Q^{(1)}_{kl}\},Q^{(2)}_{mn}\}+
\{\{Q^{(1)}_{kl},Q^{(2)}_{mn}\},Q^{(1)}_{ij}\}+
\{\{Q^{(2)}_{mn},Q^{(1)}_{ij}\},Q^{(1)}_{kl}\}= \nonumber \\
&=&(\delta_{ik}\delta_{aj}\delta_{bl}-\delta_{il}\delta_{aj}\delta_{bk}
+\delta_{jl}\delta_{ai}\delta_{bk}-\delta_{jk}\delta_{ai}\delta_{bl})
\times \nonumber \\
&\times&
[(I\circ Q^{(4)})-8\alpha(I\circ Q^{(3)})-(3-19\alpha^{2})(I\circ Q^{(2)})
+4\alpha(1-\alpha^{2})(I\circ Q^{(1)})+ \nonumber \\ 
&+&8\alpha^{2}(1-\alpha^{2})(I\circ Q^{(0)})]_{ab,mn}+ \nonumber \\
&+&(\delta_{km}\delta_{al}\delta_{bn}-\delta_{kn}\delta_{al}\delta_{bm}
+\delta_{ln}\delta_{ak}\delta_{bm}-\delta_{lm}\delta_{ak}\delta_{bn})
\times\nonumber \\
&\times&
[(I\circ Q^{(4)})-8\alpha(I\circ Q^{(3)})-(3-19\alpha^{2})(I\circ Q^{(2)})
+8\alpha(1-\alpha^{2})(I\circ Q^{(1)})+ \nonumber \\ 
&+&2(1-\alpha^{2})^{2}(I\circ Q^{(0)})]_{ab,ij}+ \nonumber \\
&+&(\delta_{mi}\delta_{an}\delta_{bj}-\delta_{mj}\delta_{an}\delta_{bi}
+\delta_{nj}\delta_{am}\delta_{bi}-\delta_{ni}\delta_{am}\delta_{bj})
\times\nonumber \\
&\times&
[(I\circ Q^{(4)})-8\alpha(I\circ Q^{(3)})-(3-19\alpha^{2})(I\circ Q^{(2)})
+8\alpha(1-\alpha^{2})(I\circ Q^{(1)})+ \nonumber \\ 
&+&2(1-\alpha^{2})^{2}(I\circ Q^{(0)})]_{ab,kl} \quad .
\end{eqnarray}
The conclusion is that the Jacobi identity also breaks beyond 
the critical point, thus for any value of $\alpha$. This is one of the main 
results of this paper.

\section{ Final remarks}

In spite of several similarities between the algebras of non-local charges 
in the $O(N)$ non-linear sigma model and the WZNW model, we have found 
a major difference: the former obeys the Jacobi identity while the later 
does not. This result came as a surprise because we expected to find 
classical Yangian algebras. Indeed, the presence of Yangians in WZNW models 
is well established \cite{22}-\cite{25} and Yangians do satisfy the 
Jacobi identity. So we looked for other indications of non-Yangian 
features: going back to those brackets in (\ref{ncritalg}), 
we noticed that the generator $Q^{(1)}$ does not transform in the 
usual manner under the $O(N)$ symmetry:
\beq
\{ Q^{(0)},Q^{(1)}\} = (I\circ Q^{(1)}) - 2\alpha 
(I\circ Q^{(0)})\quad ,
\label{29}
\eeq
except when $\alpha =0$. This implies that one of the defining relations 
of a Yangian -- the property sometimes referred to as 
Y(2) -- is not generally obeyed here. Therefore the 
cubic algebra (\ref{ncritalg}) cannot be a classical Yangian algebra. 
Neither can the algebra (\ref{critalg}) of critical charges, 
for which we nevertheless found a rather simple cubic ansatz. 

As neither the case $\alpha =0$ obeys the Jacobi identity, we have 
searched for more fundamental reasons of this violation. 
An important difference between the 
sigma model and WZNW current algebras is the presence (or absence) of 
the intertwiner field in the Schwinger terms: in the sigma 
model, the vanishing behavior of the intertwiner as $x\to \pm \infty$ 
eliminates many boundary contributions. 
In the algebra, such contributions come up as integrals of the 
type
\beq
\int dxdy\, (j\circ \partial ^{-1}j_0)(x)\, \delta '(x-y) \to 0 \quad
{\rm because} \; j(\pm \infty) \to 0 \quad ,
\eeq
where $j$ is the intertwiner field. We have used the prescription 
$\int dxdy \, F(x)\delta '(x-y) \propto [F(+\infty ) - F(-\infty )]$,
according to ref.\cite{26}.
In the WZNW model, if we consider either the left or right sector separately, 
there will be no intertwiner, and the corresponding integrals will not 
vanish, due to contributions of boundary charges:
\beq
\int dxdy\, (I\circ \partial ^{-1}J_0)(x)\, \delta '(x-y) \propto 
\left( I \circ 
(\partial ^{-1} J_0 )(+\infty) - (\partial ^{-1} J_0 )(-\infty) 
\right) = (I \circ Q^{(0)})
\eeq
For instance, the unusual second term on the rightmost side of 
eq.(\ref{29}) is originated in this way. 
Hence the algebra is bound to become more involved in the WZNW model. 
We suggest the ref.\cite{27} for a discussion on the potential breaking 
of the Jacobi identity by c-number Schwinger terms in current algebras, 
although a specific interpretation based on the WZNW-field dynamics 
would be most valuable.

As concerns the integrability of the off-critical WZNW model, 
we have not fully caught on the consequences of this intriguing 
violation of the Jacobi identity. Moreover, the algebraic study  
of this paper has not gone beyond the classical level.  
Extrapolating the examples from the non-linear sigma model, we believe 
that the commutator algebra of the infinitesimal symmetry transformations 
-- in other words, the action of the Yangian symmetry -- 
generated by the non-local charges through some Lie-Poisson action, 
should obey the Jacobi identity (see ref.\cite{16} for a discussion 
of the Lie-Poisson action of generators in the non-linear sigma model). 
In that case, the Yangian symmetry would remain associative as expected. 
Further investigations in this direction are in progress.

\vskip .5cm
\noindent {\bf Acknowledgements}

\vskip .2cm
A.Z. would like to thank the Centre de Recherches Math\'ematiques (CRM), 
Universit\'e de Montr\'eal, for its hospitality.


\begin{thebibliography}{99}

\bibitem{1}
Bernard, D.: Hidden Yangians in 2-D massive current algebras.
Commun. Math. Phys. {\bf 137} (1991) 191.

\bibitem{2}
Bernard, D.: An introduction to Yangian symmetries, in  Integrable
Quantum Field Theories, edited by L. Bonora et al., NATO ASI Series B,
Physics; vol. 310, New York, Plenum Press 1993.

\bibitem{3}
Drinfel'd, V.G.: Hopf algebras and the quantum Yang-Baxter equation.
Sov. Math. Dokl. {\bf 32} (1985) 254.

\bibitem{4}
Drinfel'd, V.G.: A new realization of Yangians and quantized affine
algebras. Sov. Math. Dokl. {\bf 36} (1988) 212.

\bibitem{5}
Leclair, A., Smirnov, F.A.: Infinite quantum group symmetry of
fields in massive 2D quantum field theory. Int. J. Mod. Phys. {\bf A7} 
(1992) 2997.

\bibitem{6}
L\"uscher, M.: Quantum non-local charges and absence of particle
production in two-dimensional non-linear $\sigma$-model. Nucl. Phys.
{\bf B135} (1978) 1.

\bibitem{7}
L\"uscher, M., Pohlmeyer, K.: Scattering of massless lumps and
non-local charges in the two-dimensional classical non-linear $\sigma$-model.
Nucl. Phys. {\bf B137} (1978) 46.

\bibitem{8}
Br\'ezin, E., Itzykson, C., Zinn-Justin, J., Zuber, J.B.: Remarks
about the existence of non-local charges in two-dimensional models. Phys. Lett.
{\bf 82B} (1979) 442.

\bibitem{9}
de Vega, H.J.: Field theories with an infinite number of conservation
laws and\break B\"acklund transformations in two dimensions.
Phys. Lett. {\bf 87B} (1979) 233.

\bibitem{10}
Abdalla, E., Forger, M., Gomes, M.: On the origin of anomalies in
the quantum non-local charge for the generalized non-linear sigma models. Nucl.
Phys. {\bf B210} (1982) 181.

\bibitem{11}
de Vega, H.J., Eichenherr, H., Maillet, J.M.: Classical and quantum
algebras of non-local charges in $\sigma$-models. Commun. Math. Phys. {\bf 92}
(1984) 507.

\bibitem{12}
de Vega, H., Eichenherr, H., Maillet, J.M.: Yang-Baxter
algebras of monodromy matrices in integrable quantum field theories. Nucl.
Phys. {\bf B240} (1984) 377.

\bibitem{13}
Maillet, J.-M.: Hamiltonian structures for integrable classical
theories from graded Kac-Moody algebras. Phys. Lett. {\bf 167B}
(1986) 401;
New integrable canonical structures in two-dimensional models. Nucl. Phys.
{\bf B269} (1986) 54.

\bibitem{14}
Mackay, N.J.: On the classical origins of Yangian symmetry in
integrable field theory. Phys. Lett. {\bf B281} (1992) 90,
erratum-ibid. {\bf B308} (1993) 444.

\bibitem{15}
Molev, A., Nazarov, M., Olshanskii, G.: Yangians and classical 
Lie algebras. CMA-MR53-93, hep-th/9409025.

\bibitem{16}
Abdalla, E., Abdalla, M.C.B., Brunelli, J.C., Zadra, A.: The Algebra of
Non-local Charges in Non-linear Sigma Models. Commun. Math. 
Phys. {\bf 166} (1994) 379.

\bibitem{17}
Saltini, L.E., Zadra, A.: Algebra of non-local charges in the 
supersymmetric non-linear sigma model. Int. J. Mod. 
Phys. {\bf A12} (1997) 419.

\bibitem{18}
Abdalla, E., Abdalla, M.C.B., Rothe, K.: Non-perturbative
methods  in two-dimen\-sional quantum field theory. Singapore: World 
Scientific 1991.

\bibitem{19}
Abdalla, E., Forger, M.: Current algebra of WZNW models at and away
from criticality. Mod. Phys. Lett. {\bf 7A} (1992) 2437.

\bibitem{20}
Mussardo, G.: Off-critical statistical models: factorized scattering
theories and bootstrap program. Phys. Rep. {\bf 218} (1992) 215.

\bibitem{21}
Abdalla, E., Abdalla, M.C.B., Sotkov, G., Stanishkov, M.: Off
critical current algebras. Int. J. Mod. Phys. {\bf A10} (1995) 1717.

\bibitem{22}
Bouwknegt, P., Ludwig, A.W.W., Schoutens, K.: Affine and Yangian 
Symmetries in $SU(2)_1$ Conformal Field Theory. hep-th/9412199.

\bibitem{23}
Ahn, C., Nam, S.: Yangian Symmetries in the $SU(N)_1$ WZW Model and 
the Calogero-Sutherland Model. Phys. Lett. {\bf B378} (1996) 107.

\bibitem{24}
Bouwknegt, P., Schoutens, K.: The $SU(n)_1$ WZW Models: Spinon 
Decomposition and Yangian Structure. Nucl. Phys. {\bf B482} (1996) 345.

\bibitem{25}
Bernard, D., Maassarani, Z., Mathieu, P.: Logarithmic Yangians in 
WZW models. Mod. Phys. Lett. {\bf A12} (1997) 535.

\bibitem{26} 
Gelfand, I.M., Chilov, G.E.: Les Distributions. Ed. Dunod, Paris, 
1962.

\bibitem{27} 
Treiman, S.B., Jackiw, R., Gross, D.J.: Field Theoretic Investigations 
in Current Algebra, in Lectures on Current Algebra and Its Applications. 
Princeton University Press, NJ, 1972. 

\end{thebibliography}
\end{document}